# From the field

## Peer-review Platform for Astronomy Education Activities

**Authors**

**Pedro Russo**
russo@strw.leidenuniv.nl
astroEDU Managing Editor

**Thilina Heenatigala**
heenatigala@strw.leidenuniv.nl
astroEDU Assistant Editor

Leiden Observatory / Leiden University
The Netherlands

**Edward Gomez**
egomez@lcogt.net
astroEDU Managing Editor
Las Cumbres Observatory Global Telescope Network (LCOGT)
California, USA

**Linda Strubbe**
linda@cita.utoronto.ca
astroEDU Editor in Chief
Canadian Institute for Theoretical Astrophysics
Toronto, Canada

**Tags**

Astronomy Education, Open Educational Resources, Web technologies, Educational Repositories

Hundreds of thousands of astronomy education activities exist, but their discoverability and quality is highly variable. The web platform for astronomy education activities, astroEDU, presented in this paper tries to solve these issues. Using the familiar peer-review workflow of scientific publications, astroEDU is improving standards of quality, visibility and accessibility, while providing credibility to these astronomy education activities. astroEDU targets activity guides, tutorials and other educational activities in the area of astronomy education, prepared by teachers, educators and other education specialists. Each of the astroEDU activities is peer-reviewed by an educator as well as an astronomer to ensure a high standard in terms of scientific content and educational value. All reviewed materials are then stored in a free open online database, enabling broad distribution in a range of different formats. In this way astroEDU is not another web repository for educational resources but a mechanism for peer-reviewing and publishing high-quality astronomy education activities in an open access way. This paper will provide an account on the implementation and first findings of the use of astroEDU.

## 1. Introduction

The amount of educational content freely available on the Web is large and growing fast. Many challenges have emerged for educators when looking for and comparing resources available online, most of them related with discoverability, quality and openness of the resources. The Open Educational Resources (OERs) model (Hylén, 2006) addressed some of these challenges, offering a new, scalable, and potentially powerful vision of learning. OERs are teaching, learning and research resources that reside in the public domain or have been released under an intellectual property license that permits their free use or re-purposing (Atkins et al., 2007). This concept has been further developed: making OERs be cost free to the end-user; allowing the end-user freedom to Reuse, Revise/alter, Remix and Redistribute, the 4R framework. This framework was initially presented by Wiley and in expanded in detail by Hilton III et al. (2010) (Figure 1).

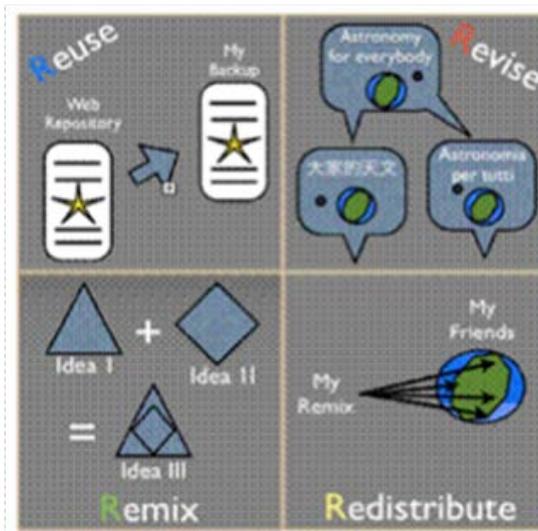

Figure 1. Open Education Resources 4R framework: Reuse, Revise(Alter), Remix, Redistribute as presented by Hilton III, Wiley, Stein & Johnson (2010)

The far-reaching impact of the OERs in society is not widely understood and is full of challenges on creation, use and evaluation (Smith & Casserly, 2006). Not all of these challenges derive from the OER model, but from lateral reasons, for example the use of OER is connected with the level of Internet access, knowing that two thirds of the world's population still doesn't have Internet access or the educational policies at different levels, within institutions and in government (Morley, 2012). Moreover OER is seen as a potential threat to education content held by publishing houses (OECD, 2007).

Nevertheless the number of OER has opened a new way for science education to produce, develop and distribute resources. The number of repositories that store these resources has been growing in recent years, each with a different emphasis. Below we present a summary of repositories which are specific to astronomy education activities and resources.

## 2. Science Education Repositories

Although there are thousands of educational repositories [10], archiving a variety of resource types, there are not many repositories of educational resources specifically for astronomy. The table below gives an overview of existing repositories for astronomy education in English.

| Repository | Compadre (Physics and Astronomy Education Activities) | Galileo Teachers Training Program | Astronomical Society of the Pacific | NASA Wavelength | astroEDU |
| --- | --- | --- | --- | --- | --- |
| URL | www.compadre.org | www.galileoteachers.org | www.astrosociety.org/education/ | www.nasawavelength.org/ | www.iau.org/astroEDU |
| Type of review | Internal review | Internal review | Internal review | Internal review | Internal review |
| Open to submissions? | Yes | No | No | No | Yes |
| Multilingual Support? | No | No | No | No | Planned |
| User registration required? | No | No | Yes | No | No |
| Library Type | Repository (Resource records) | Repository (Resources in PDF) | Listing (lists of links to resources) | Repository (Resource records) | Library |
| License | Various (from no license to copyrighted) | Various (from no license to copyrighted) | Various (from no license to copyrighted) | Various (from no license to copyrighted) | Creative Commons Attribution 3.0 Unported |
| Resource Types | Student resources, Teaching resources, Games, A/V material, Tools, Datasets, Reference materials | Teaching resources, Tools, webpages. | Reference materials, Courses, Webpages | Student resources, Teaching resources, Datasets, Reference materials, News & events, Webpages | Lessons Plans. |
| Connection to the Curriculum | None | None | None | US National Science Education Standards | Yes (English version: UK, Australia, US) |
| Collection growing? | N/A | N/A | N/A | Yes | Yes |
| Reference | Deustua, 2004 | N/A | N/A | Smith, 2013 | Russo, Gomez, et al. 2014 |

Table 1.

The information collated on those repositories is mostly well organized and of high quality, however none of those repositories satisfies The information collated on those repositories is mostly well organized and of high quality, however none of those repositories satisfies completely the OER 4R model previously mentioned; sometimes the material cannot be revised or remixed; it can only, in most of the cases, be redistributed and reused. All of these repositories are only available in English and the majority do not provide the original source text of materials, which would facilitate adaptation and translation, essential for an international platform.

The discoverability of resources is also one of the problems. There is no evidence that teachers use the repositories of educational material to find their resources. Some educators use generic search engines, like Google. Results from a Google search provide very little indication of the quality of the resources. However quality is one of the most important criteria for educators when they search for learning resources online (Brent, 2012).

## 3. astroEDU

Although OERs offer a good solution for sharing knowledge, particularly when putting open educational resources on the internet, ensuring these OERs are also of high quality remains a challenge. To address this we propose astroEDU (www.iau.org/astroedu) an online platform for sharing astronomy OERs. astroEDU conforms to the 4R framework but which adds a new component, review, to ensure the resources are of the highest quality. In this respect astroEDU has enhanced the 4R model to a 5R model where Review becomes the fifth 'R'. To address the need to review the resources we propose a review system similar to that used in the academic knowledge creation and dissemination; the peer-review model.

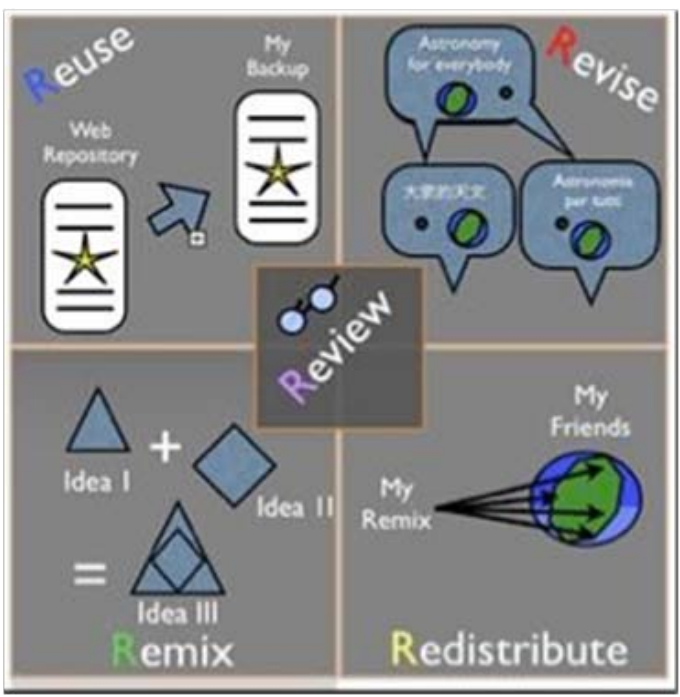

Figure 2. Proposed new OER framework: 5R: addition of Review of content through scientific and pedagogical quality checked (and improved) by the community peers

Peer review was introduced to scholarly publication more than 280 years ago, as a process to evaluate any work by one or more people of similar competence to the producers of the work (also called peers). In academia this constitutes a form of self-regulation by qualified members of a profession within the relevant field. Peer review methods are employed to maintain standards of quality, improve performance, and provide credibility. In academia, peer review is often used to determine an academic paper's suitability for publication (Scanlan 2013). Interestingly the peer-review process hasn't been widely used for the evaluation of science education products such as educational activities. Early attempts of using this methodology have been done by Cafolla (2006) and Gold (2012) with varying levels of success.

More recent attempts have been made by NASA Wavelength and Climate Literacy and Energy Awareness Network (CLEAN) to integrate peer-selection and internal peer-review. The resources are selected by a panel or by team members and then reviewed by a pre-appointed review board. These review methodologies enhance the quality of the resources, but do not provide a system for any educator to submit their resources, which is a limitation of both platforms.

The innovative aspect of astroEDU is the use of peer-review in a similar way to its use in scholarly publication. The suitability for publication of the activity is evaluated by the two stakeholders peers; an educator and an astronomy researcher. In this way both educational and scientific accuracy of the activity is checked and reviewed.

For astroEDU the methodology used is anonymous peer review, also called blind review. In this system of pre-publication peer review of scientific articles or papers for journals by reviewers who are known to the journal editor but whose names are not given to the article's author. The reviewers do not know the author's identity, as any identifying information is stripped from the document before review. In this respect astroEDU's form of peer-review is double blind and free from bias. Moreover the same way that peer-reviewed scholarly articles are the main metruc for performance evaluation of scholars, astroEDU will provide a new metric to assess the quality of the work developed by educators.

To ensure a rigorous peer-review of educational activities an activity template was established and designed by the astroEDU editorial board. For that specific learning outcomes need to be identified, which enable the logic of the activity and the evaluation structure. For astroEDU educational taxonomy was established based on (AGU, 2013). astroEDU activities follow the standard components defined by several authors in science education (USU 2008). The components can be broken in four main areas: Objectives (What will your students learn?), Materials (What are the teaching instruments?), Processes (How will you teach your students?) and assessment (How will you assess your students' learning?). In Table 2 you can find a detailed explanation for the relevant different sections of astroEDU activities.

| Sections of astroEDU activities. | Description |
| --- | --- |
| Activity title | Full title of the activity. |
| Keywords | Any words that relate to the subject, goals or audience of the activity. Note that most submissions will get variety of keywords and editor must ensure to select and add relevant keywords. Important for on-line search. |
| Age range | All age categories the activity applies to. The categories may change depending on the reviewers' and editorial board's input |
| Education level | The education level will change depending on the reviewers' and editorial board's input. |
| Time | The time taken to complete the activity. |
| Group size | Defines whether the activity is for individual or group use. Can also provide information like how many students per teacher |
| Supervised for safety | Determine whether the activity has steps that require adult supervision for safety. E.g.: using scissors. |
| Cost | Estimated cost of any materials needed for the activity. For astroEDU as a currency we use Euro (€). |
| Location | Suitable location to conduct the activity (for example indoors or outdoors). |
| List of material | List of items needed for the activity. Try to find materials which are easily and cheaply available in most countries (or offer alternatives) |
| Overall Activity Goals | A short list of points outlining the general purpose of the activity, and why these are important for students to learn. For example, "The overall goals of the activity are for students to understand why we experience seasons, and to improve their ability to communicate scientific findings. Seasons are important to understand because they affect daily life and involve concepts about light that are important in other contexts as well." (More specific learning objectives are entered in the field "Learning Objectives") |
| Learning Objectives | Learning objectives are specific statements that define the expected goals of an activity in terms of demonstrable skills or knowledge that will be acquired by a student as a result of instruction. These are also known as: instructional objectives, learning outcomes, learning goals. The demonstration of learning should be tied directly to "Evaluation". On the following page you can find some additional information on how to formulate good learning objectives: http://edutechwiki.unige.ch/en/Learning_objective. Use terminology listed on the page. For example, "Students will use the concept of solar flux as a function of incidence angle to explain why it is hot in summer and cold in winter in Toronto." |
| Evaluation | Include ways to test the goals, learning objectives and key skills learned by the audience. A way to assess the implementation of the activity and your performance should also be included. |
| Background information | This section contains information that teachers will read prior to beginning the activity. Necessary scientific background information needed to implement this activity. Limit each topic to one paragraph and keep in mind what is actually necessary for the activity. Also keep in mind the background of the teacher (e.g., explain concepts clearly, and do not use inappropriately technical vocabulary). |
| Core skills | Determine whether the activity is; Asking questions, Developing and using models, Planning and carrying out investigations, Analysing and interpreting data, Using mathematics and computational thinking, Constructing explanations, Engaging in argument from evidence, Communicating information or a combination of these. |

| Type of learning activity | Enquiry models are a pedagogical methodology for learning activities where the educational activity "starts by posing questions, problems or scenarios, rather than simply presenting established facts or portraying a smooth path to knowledge". There are several approaches to enquiry-based instruction. These approaches include; Open-ended enquiry, Guided enquiry, Structured enquiry, Confirmation or Verification, Fun Learning. |
|---|---|
| Brief Summary | One-paragraph short description of the activity. The description should give an introduction to the activity as well as what to expect from completing the activity. |
| Full description of the activity: | Detailed step-by-step breakdown of the activity. Use graphics where possible to show the steps. |
| Connection to school curriculum | Add the curriculum connection from the relevant country or region. The astroEDU editorial board will help find further connections. |

Table 2. astroEDU Educational Activities Taxonomy

## 4. astroEDU Technical Implementation

The publication workflow of astroEDU was designed to remove barriers to the creation, submission use and re-use, and sharing of high-quality content. To achieve these goals, astroEDU uses off-the-shelf web technologies for the production and publication workflows.

Submission is done via e-mail or a web form (typeform). Google documents and spreadsheets are used as collaborative tools within the editorial workflow, as described in Figure 3. Central to the philosophy of astroEDU is disseminating the best astronomy education resources. The majority of educator interaction with astroEDU will be searching and browsing for resources on the website. After review, activities are made available in many different formats: PDF, .doc, HTML, and epub, including the source files (RTF) for future translations and remixes. These successful activities are then syndicated through educational resources repositories and sharing sites (example: Scientix (ref), TES (ref.) and OER.). One of the main goals of the astroEDU is to promote the use of excellent activities worldwide. That is the reason why all the astroEDU activities will be licensed through the Creative Commons Attribution 3.0 Unported license. All the astroEDU activities are labeled with a Digital Object Identifier (DOI), to provide a form of persistent identification for future reference and an easy way to educators to reference their activities just like in scholarly paper .

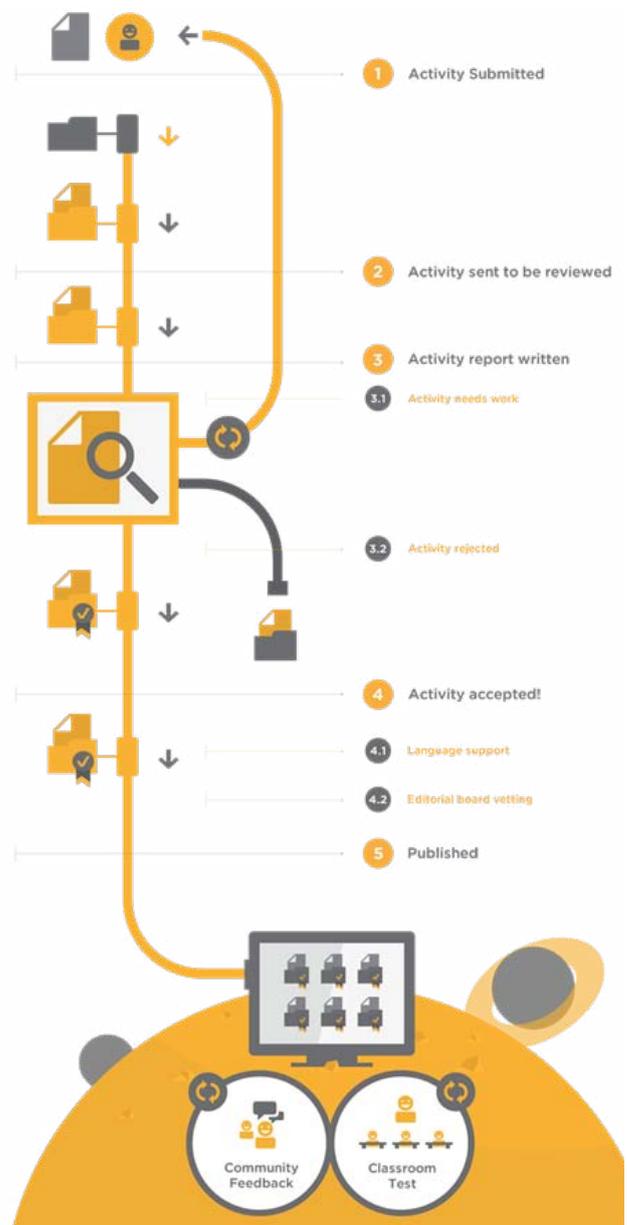

Figure 3. Production pipeline (to be produced)

The front-end website uses different web technologies, mainly open-source software. The table 3. gives you an overview of the different web technologies.

| Web Technology | Use in astroEDU | Reference |
| --- | --- | --- |
| Django | Django is a high-level Python Web framework that encourages rapid development and clean, pragmatic design. Django was designed to handle two challenges: the intensive deadlines of a newsroom and the stringent requirements of the experienced Web developers who wrote it. It lets you build high-performing, elegant Web applications quickly. | Forcier, J., Bissex, P., & Chun, W. (2009). Python Web development with Django. Upper Saddle River, NJ: Addison-Wesley. |
| Python | Python is a dynamic object-oriented programming language that can be used for many kinds of software development. It offers strong support for integration with other languages and tools, comes with extensive standard libraries, and can be learned in a few days. | Forcier, J., Bissex, P., & Chun, W. (2009). Python Web development with Django. Upper Saddle River, NJ: Addison-Wesley. |
| MariaDB | MariaDB is a robust, scalable, and reliable SQL server. MariaDB is a drop-in replacement for MySQL. | MariaDB An enhanced, drop-in replacement for MySQL. (n.d.). Retrieved May 22, 2014, from https://mariadb.org/en/ |
| NGINX | NGINX is a high performance, open source web application accelerator that helps websites deliver more content, faster, to its users. | Nginx news. (n.d.). Retrieved May 22, 2014, from http://nginx.org/ |
| Memcached | Memcached is a high-performance, distributed memory object caching system, generic in nature, but intended for use in speeding up dynamic web applications by alleviating database load. | Nginx news. (n.d.). Retrieved May 22, 2014, from http://nginx.org/ |
| Elasticsearch | Elasticsearch is a search server based on Lucene. It provides a distributed, multitenant-capable full-text search engine with a RESTful web interface and schema-free JSON documents. | Elasticsearch. (n.d.). Retrieved May 22, 2014, from http://www.elasticsearch.org/overview/ |
| Markdown and ReportLab | PDF and EPUB are digital formats optimised for printing and e-books, respectively. AstroEDU activities are available in these formats to enable broader use, leveraging technologies such as Markdown and ReportLab. | Markdown: Syntax. (n.d.). Retrieved May 22, 2014, from http://daringfireball.net/projects/markdown/syntaxhttp://daringfireball.net/projects/markdown/syntax http://daringfireball.net/projects/markdown/syntax ReportLab open-source PDF Toolkit. (n.d.). Retrieved May 22, 2014, from http://www.reportlab.com/opensource |

Table 3. Web technologies used to develop and implement astroEDU.

## 5. Conclusions & Future Work

astroEDU is an open-access platform for peer-reviewed astronomy education activities and makes the best astronomy activities accessible to educators around the world. As a platform for educators to discover, review, distribute, improve, and remix educational astronomy activities, astroEDU tries to solve some past issues with educational activities in astronomy, namely that high-quality educational activities are difficult to find. As a proof-of-concept, the whole astroEDU process demonstrates that through peer-review and publication in a simple, elegant website, high-quality educational materials can be made available in an open-access way. The initial results and feedback is positive: "[AstroEDU is] off to a promising start, with a pleasing range of activities suited to children of all ages and abilities" (Physics World, 2014).

The pedagogical impact of astroEDU will be measured in the next years, when more activities will populate the repository and more educators will use the materials. In the near future astroEDU will also explore new ways to review its content, mainly through classroom evaluation and post-publication evaluation. For classroom evaluation some randomized evaluations will be run in schools in Wales (UK) and the Netherlands. The different educators can also use the comments box for each activity so who use the activity can discuss how it worked when they used it, testing, etc. astroEDU will also test new models of peer-review in contrast with the current anonymous peer-review, some existing models like open peer-review will be tested. The discoverability of educational material is another issue that will be addressed in the next developments steps. Using techniques like Search Engine Optimization we expect to increase the number of users for the astroEDU activities. astroEDU is currently available in English, although astroEDU currently welcomes submissions in any language. It is anticipated the platform will be offered in other languages in early 2015. Only a truly cross platform and cross-language experience will be useful for educators and teachers around the world and astroEDU will try to achieve that.

## Acknowledgement:

astroEDU was developed by funding from the European Community's Seventh Framework Programme ([FP7/2007-2013]) under grant agreement n° 263325. astroEDU is a project of the International Astronomical Union's Office of Astronomy for Development. astroEDU development was supported by, International Astronomy Union, Universe Awareness, Leiden University, LCOGT and European Union. We would like to thank Silvia Simionato, Alejandro Cárdenas-Avendaño, Bruno Rino and Jan Pomierny for their comments to this article.

# From the field